\documentclass[prl,aps,twocolumn,showpacs]{revtex4}

\def\iT1T{$(T_1T)^{-1}$}
\def\iT1{$T_1^{-1}$}

\usepackage{epsfig}

\begin{document}

\title{Phase Inhomogeneity of the Itinerant Ferromagnet MnSi at High Pressures}
\author{W. Yu$^{1}$, F. Zamborszky$^{1,2}$, J. D. Thompson$^{2}$, J. L.
Sarrao$^{2}$, M. E. Torelli$^{3}$, Z. Fisk$^{3}$ and S. E. Brown$^{1}$}
\affiliation{$^1$Department of Physics and Astronomy, University of
California at Los Angeles, Los Angeles, CA 90095-1547}
\affiliation{$^2$Los Alamos National Laboratory, Los Alamos, NM 87545}
\affiliation{$^3$National High Magnetic Field Laboratory, Florida State University,
Tallahassee, FL 32310} 

\date{\today}

\begin{abstract}
The pressure induced quantum phase transition of the weakly itinerant
ferromagnet MnSi is studied using zero-field $^{29}Si$ NMR spectroscopy
and relaxation. Below $P^*\approx 1.2GPa$, the intensity of the signal
and the nuclear spin-lattice relaxation is independent of pressure, even
though the amplitude of the magnetization drops by 20\% from
the ambient pressure amplitude. For $P>P^*$, the decreasing intensity
within the experimentally detectable bandwidth signals the onset of an
inhomogeneous phase that persists to the highest pressure measured,
$P\ge 1.75GPa$, which is well beyond the known critical pressure
$P_c=1.46GPa$. Implications for the non-Fermi Liquid behavior observed for $P>P_c$ are
discussed.
\end{abstract}
\pacs{71.10.Hf, 75.50.Cc, 76.60.-k}
\maketitle

In magnetically-ordered systems where the magnetism is weak, the
possibility of tuning the
transition temperature $T_c\to0$ using an adjustable external parameter
is of particular importance, because in the case that the transition is
continuous, the Fermi Liquid description is expected to break down at
the quantum critical point (QCP) \cite{Hertz-Millis}. Further, there are
many systems known where the physical properties cannot be described as
Fermi Liquids and so far, it is debated to what extent this can be
accounted for by the proximity to a QCP. Recently, the ability to tune
the magnetic ordering transition in ostensibly clean systems using high
pressure was demonstrated for several itinerant ferromagnetic systems,
including $MnSi$ \cite{Thompson,Pfleiderer1}, $ZrZn_2$ \cite{Pfleiderer3},
and $UGe_2$ \cite{Saxena}, as well as the antiferromagnet
heavy-fermion systems $CeIn_3$, $CePd_2Si_2$ \cite{Mathur}, as well as several in the
$CeMIn_5$ family \cite{Sidorov}. In many of these examples, non-Fermi Liquid behavior
is observed over a range of temperatures and pressures, although
applying theory to the experiments is not straight-forward for
either case \cite{Coleman,Pfleiderer2}.

Consider the weak itinerant ferromagnetism in MnSi, the most
extensively studied of the weak ferromagnets. Its ambient-pressure
properties are successfully described by the self-consistent
renormalization (SCR) theory of spin fluctuations \cite{Moriya}. First,
it is characterized by a Curie-Weiss (CW) form for the susceptibility,
where the effective moments $\mu _{eff}(T>T_c)$ are much larger than the
saturation moments $\mu _s(T\ll T_c)$ \cite{Wernick}. The low energy
excitations probed by neutron scattering \cite{Ihsikawa2} and the
critical behavior probed by NMR \cite{Yasuoka} and $\mu SR$ \cite{musr}
are consistent with exchange-enhanced spin fluctuations \cite{Moriya}.
Below $T_C$, MnSi orders as a helimagnet with a very small wavevector ,
$Q\approx 2\pi (1,1,1)/180A^{\circ}$ \cite{Ihsikawa}, due to the
Dzyaloshinski-Moriya (DM) interaction arising from spin-orbit coupling
in the $B20$ lattice structure lacking inversion symmetry \cite{Bak}.
The ground state is a Fermi Liquid \cite{Pfleiderer1}. With applied
pressure, the transition temperature smoothly approaches zero
temperature \cite{Thompson}. In approaching the critical pressure and
beyond, there are several observations that fall outside the
Hertz-Millis framework. First, at pressures starting from $1.2Gpa$ and
up to the critical pressure $P_C\approx 1.46GPa$, the transition is
first order \cite{Pfleiderer1}.
Next, it is known empirically that for a remarkably wide range of
pressures $P>P_c$, the temperature-dependent part of the resistivity
varies as $\Delta\rho\propto T^{3/2}$ to temperatures well below where
the crossover to Fermi Liquid behavior is expected
\cite{Doiron-Leyraud}. It appears to be a general phenomenon, as similar
behavior is reportedly observed for $ZrZn_2$ and Ni$_{3}$Al
\cite{Pfleiderer2}. In the specific case of MnSi, for $P>1.2GPa$, the
relationship of the broad maximum in ac susceptibility
\cite{Pfleiderer1} in the vicinity of the critical pressure and the
non-Fermi Liquid behavior is unclear.
 These observations led Pleiderer, {\it et al.}
\cite{Pfleiderer2}, to question whether the ground state in nearly
ferromagnetic metals is a Fermi Liquid at all.
In addressing this question, NMR
and $\mu^+$SR ought to be a valuable probe of the local
physics.

In this Letter, zero-field $^{29}$Si NMR spectroscopy and dynamics on
powdered samples of MnSi are presented, with particular emphasis on
pressures in the vicinity of $P_c$. Due to the homogeneous and isotropic
hyperfine coupling constant $A_{hf}=58.7kOe/\mu B$ \cite{Motoya},
the NMR resonance from $^{29}$Si gives approximately the local moment,
about $0.4\mu _B/Mn$ at ambient pressure\cite{Wernick}. In monitoring the low
temperature pressure and paying particular attention to quantifying
pressure inhomogeneities, we find the spin dynamics, as inferred from
$(T_1T)^{-1}$, undergoes a discontinuous change when the pressure exceeds
$P^*\approx 1.2-1.3GPa$. Coincident with the change in dynamics is a
smooth decrease in signal intensity for $P>P^*$. We find that the signal persists to the
highest pressure we were able to measure,
$P\approx 1.75GPa$, far beyond the quoted critical pressure
$P_c=1.46GPa$ \cite{Pfleiderer1}. The observations are consistent with a change in the
magnetic state for pressures exceeding $P^*$, where the system also
becomes inhomogeneous. Perhaps the unusual transport behavior seen for
pressures exceeding $P^*$ can be attributed to the magnetic
inhomogeneity. We make the important
acknowledgment that there are conflicting reports in the literature
regarding the existence of a magnetic phase for $P>P_c$ in MnSi
\cite{Thessieu}. In the results reported here, we have kept track of the
mean pressure and assign an upper bound to the extent of pressure
inhomogeneities. We accounted for the frequency dependence of our
sensitivity and rf enhancement factors in normalizing the signal
intensities.

The MnSi samples were made by a rf induction melting technique. The
materials were ground to a grain size $d\approx 10\mu m$ in order to
achieve significant bulk rf penetration. At this size, we found
rf enhancement factors are independent of rf power levels. The
transverse spin relaxation $T_2$ was measured by the spin echo
technique, while the longitudinal spin relaxation rate \iT1\ was
measured by the inversion-recovery method, with proper phase cycling to
remove any stimulated echoes from the accumulated transients. Perhaps
because the ground state is a helimagnet, no evidence for domain wall
contribution was found in the powdered samples as one might find for
ferromagnets \cite{Weger}. The applied pressure at low temperature was
calibrated by measuring the $^{63}Cu$ NQR frequencies in $Cu_2O$ powder
\cite{Reyes} mixed with the MnSi and placed in the same coil. The mean
pressure could be established to within 0.01 GPa, and an upper bound for
the pressure inhomogeneities was inferred from the linewidth of the
$^{63}Cu$ NQR signal.

\begin{figure}
\includegraphics[width=7cm, height=7cm]{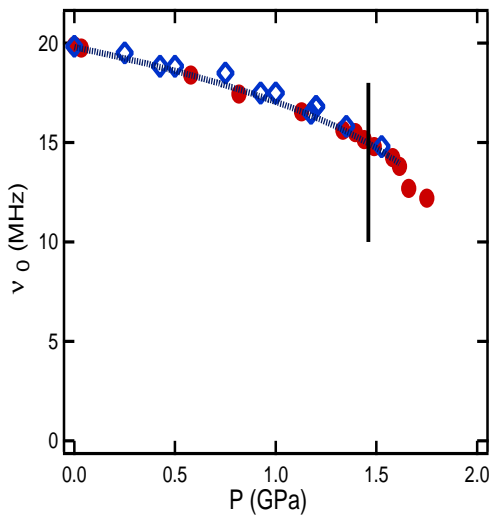}
\caption{\label{freq} Resonant frequency $\nu _0$ of the NMR absorption
 at different pressures (solid circles). Dashed line is a guide to the eye. 
Diamonds represent the saturation magnetization data from Ref. \cite{Koyama}. }
\end{figure}
The resonant frequency vs. pressure recorded at $T=1.8K$ is shown in
Fig. \ref{freq}. The values are taken to be the frequency of maximum NMR
signal, $\nu _{0}(P,T=1.8K)$ for swept frequency; we will return to this
point below, as it relates to the pressure inhomogeneity.
Our measurements at low temperatures show that $\nu_{0}$ almost
saturates at this temperature, and the overall variation is
quantitatively similar to previous reports \cite{Thessieu}. With
increasing pressure, $\nu_{0}$ decreases slowly up to $1.46GPa$. This
pressure, as indicated by the vertical line in the figure, is the
critical pressure $P_c$ where ac susceptibility measurements indicate
$T_c\to 0$ \cite{Pfleiderer1}. Surprisingly, an NMR signal is still
observed up to $1.75GPa$, accompanied by a fast drop of resonance
frequency and intensity. Although the spectrum is further
inhomogeneously broadened with increasing pressure, the peak feature
remains well-defined to highest pressures.

Before presenting the relaxation rates, we would first like to address
the experimental significance of the high-pressure NMR signal. As there
is no applied field, its presence indicates local static magnetism, on
NMR time scales, for $P>P_c$. While it is compelling to attribute this
observation to pressure inhomogeneities, the linewidths of the $^{63}Cu$
NQR signal are too narrow to support that interpretation. An indication
that our results for the pressure dependence of the static moments are
consistent with independent measurements comes from a
comparison of the bulk saturation moments \cite{Koyama} and $\nu_{0}$
over a range of pressures. The two results are shown in Fig. \ref{freq};
the moments are normalized to the NMR frequencies using {\it only}
$A_{hf}$.

\begin{figure}
\includegraphics[width=7cm]{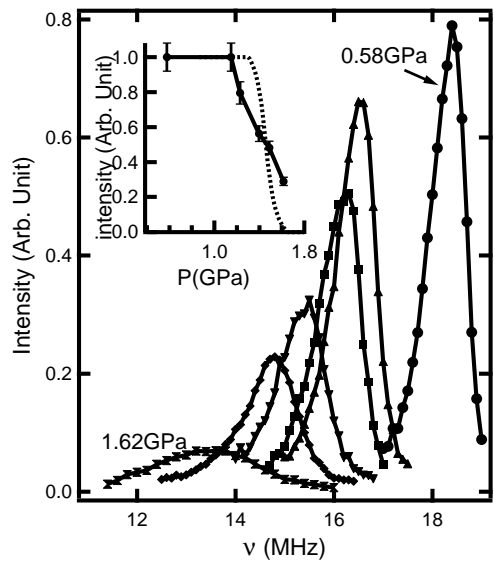}
\caption{\label{lineshape}Normalized $^{29}$Si spectrum at six pressures, from
right to left, $p=0.58$, $1.15$, $1.23$, $1.40$, $1.49$, and
$1.62GPa$. Inset: Comparison of total spectral intensity (solid
circles) and the expected one (dashed line) due to pressure inhomogeneity,
 assuming a first order phase transition at $P_c=1.46
GPa$.}
\end{figure}
The $^{29}$Si spectrum at different frequencies are scaled to correct
for temperature- and frequency-dependent sensitivities, as well as for
the rf enhancement of the magnetic phase. Figure \ref{lineshape} shows
the normalized intensity at six different pressures. In addition to the
decrease of the central resonance frequency $\nu _0$ with pressure, the
spectra are inhomogeneously broadened at all pressures. Sample and
pressure inhomogeneities both contribute to the line broadening. The
pressure inhomogeneity that occurs in the solidified pressure medium
(Flourinert Fc-75) is determined from the temperature-dependent
linewidth of $^{63}$Cu NQR spectrum. At $T=1.8K$, the spectrum is near
to Lorentzian shape, and the resulting pressure inhomogeneity gives
$\Delta P/P\approx 4\%$, that is, $\Delta P\approx 0.06GPa$ at
$P=1.5GPa$. In addition, the $^{29}$Si spectrum at $P=0.58GPa$ is
noticeably asymmetric, with the weight of the low frequency tail larger
than the high frequency side. This indicates the phase is inhomogeneous,
although we are not sure if this is intrinsic or due to quenched
disorder. The integrated spectral intensity at different pressures are
plotted in the inset of fig.~\ref{lineshape}. An unexpected intensity
loss is seen from $1.2GPa$, which is labeled as $P^*$.  For comparison,
the expected spectrum due to pressure inhomogeneity, assuming a first
order phase transition at $P_c=1.46GPa$, is also plotted in the inset of
Fig.~\ref{lineshape}. The range of pressures over which we observe the
decrease of the signal is much wider than we would predict from the
upper bound on the pressure variation alone.

\begin{figure}
\includegraphics[width=7cm]{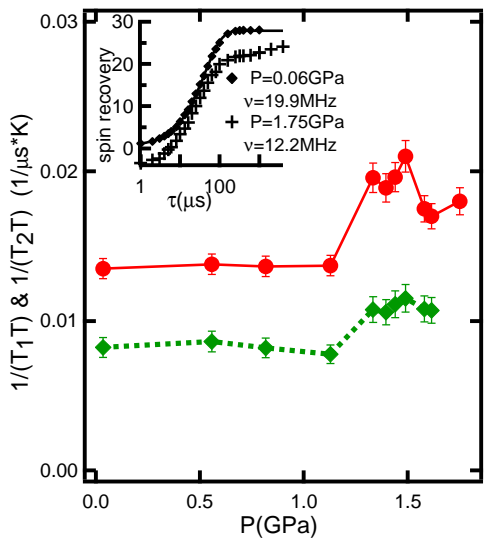}
\caption{\label{relax} Spin-lattice relaxation rate $T_1^{-1}(\nu _0)$ 
(circles) and spin-spin relaxation rate $T_2^{-1}(\nu _0)$ (diamonds) at different pressures.
In the inset is a comparison of longitudinal magnetization recovery for 
a pressure $P<P^{*}$, and $P=1.75GPa$. The amplitudes of the recoveries are adjusted
for clarity, and the solid line is a single-exponential recovery.}
\end{figure} 
The relaxation rates, for longitudinal and transverse relaxation, are
plotted as a function of pressure in Fig.~\ref{relax}.  For fixed
pressure, transverse spin relaxation follows the relation $\nu
_0^2/T_1T=constant$ at low temperatures, as expected from spin
fluctuation theory \cite{Moriya2} and previously observed \cite{musr, Thessieu}.
However, For pressures below $P^*$,
$1/T_1T$ is independent of pressure even though $\nu _0$ dropped by
$25\%$. We also observed that $T_2$ is proportional to $T_1$, with
$T_2/T_1\approx 5/3$. Assuming that the spin-lattice relaxation is
dominated by fluctuations in the hyperfine field, we expect that
$T_2=T_1$ when the same fluctuations dominate the spin echo decay and
they are isotropic. The anisotropy of the ordered phase leads to much
larger field fluctuations in the direction transverse to the average
spin orientation, and the unusual result $T_2>T_1$ \cite{Slichter}.
Above $P^*$, however, both $1/T_1T$ and $1/T_2T$ increases dramatically
and peaks near to $P_c$. This observation, together with the signal loss
fo4 $P>P^*$, suggests that the high-pressure phase has different properties
from the long-range ordered helical phase observed at lower pressures.
In addition, the longitudinal spin recovery for both below and above $P^{*}$
are adjusted and compared in the inset of Fig.~\ref{relax}. In contrast to the 
behavior at low pressures $P<P*$ where the recovery is always described by a 
single exponential, the non-exponential recovery observed when $P>P^{*}$ demonstrates 
that there are quite different environments and further work is necessary to clarify the details.

\begin{figure}
\includegraphics[width=7cm, height=6cm]{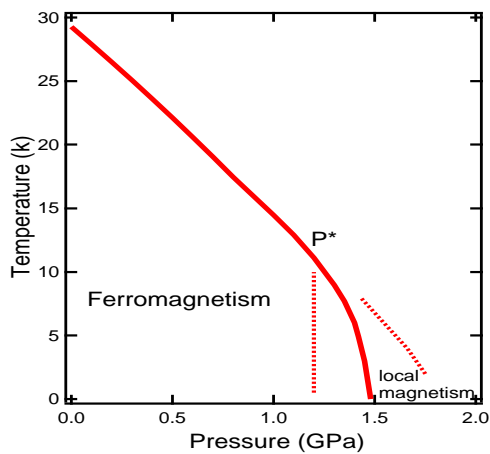}
\caption{\label{diagram}A phase diagram after \cite{Pfleiderer1}
(solid line) and an onset of inhomogeneous phase between dotted lines from
present NMR measurements. }
\end{figure}
For comparison, the ferromagnet transition temperature after Ref.
\cite{Pfleiderer1} is plotted as the solid line in Fig.~\ref{diagram}.
From present NMR measurements, we propose inserting another: The nearly
vertical line at $P^*$ is our estimate for the onset of phase
inhomogeneity.
The other is merely demarking the range of temperature and pressure
where local magnetism was detected. That is, where sites with static moment
$\mu\ge 0.2\mu_B$. Due to the reduction of the signal intensity and tank
circuit sensitivity, it became impractical to quantify the signal
strength below $\approx 10MHz$, although there is no reason to suspect that there
are not $^{29}Si$ nuclei with correspondingly low Larmor frequencies,
particularly at higher pressures.

It is compelling to cast these results in the context of a model
recently advanced to explain high-pressure resistivity measurements on very
clean, single crystal MnSi \cite{Pfleiderer2, Doiron-Leyraud}. In zero applied magnetic field,
the principle observation is that to the lowest measured temperatures, the
temperature-dependent part of the resistivity $\Delta\rho\cong T^2$ for
$P<P_c$, and $\Delta\rho\cong T^{3/2}$ for $P>P_c$. Remarkably, the
non-Fermi Liquid exponent persists at least to the maximum reported pressure
tested, namely $P=2.75GPa$. The authors propose that droplet-like phase
inhomogeneity occurs as a result of proximity to the tricritical point. Under some
conditions, such a model would lead an NMR signal as observed. However,
the inhomogeneities observed in the $^{29}Si$ NMR signals over such a 
wide range of pressures about $P_c$ could also point to the importance of static
disorder in the vicinity of the quantum phase transition. On the other hand, 
non-magnetic quenched disorder is expected to suppress the first order 
transition and inhomogeneities that result \cite{Belitz}.
Perhaps competing interactions are relevant here, as they are known to be
important in MnSi: for example, $Mn_{.9}Rh_{.1}Si$ is a spin glass \cite{Yu}.
Further exploration of dynamics associated with the NMR signals, in principle,
could give further justification to models based on fluctuations of $P_c$ in
clean systems.

To summarize, $^{29}Si$ NMR spectroscopy and dynamics in MnSi under
pressure demonstrate the existence of static magnetism persisting to higher
pressures than previously recognized. In particular, we observe a zero-field
$^{29}Si$ NMR signal at $1.75GPa$ even though the cited critical pressure is
$P_c\approx 1.46GPa$. We calibrated the pressure using the $^{63}Cu$ NQR
line in $Cu_2O$, and used it to rule out pressure variations as the reason
for our observations. There is a large pressure dependence to the signal
intensity starting from $P^*=1.2GPa$. It may be fortuitous, but this is precisely
the pressure where the magnetic ordering transition changes from second
order to first order. The change in the spin dynamics, which we infer from the
larger value of $(T_1T)^{-1}$ indicates that the inhomogeneous magnetism is locally
different from the helimagnet at low pressures. NMR experiments in a
magnetic field could be used to determine whether the higher pressure magnetism
remains helical. Although we are unaware how much disorder plays a
role in this phase, the unexpected properties observed in this class of itinerant
ferromagnet near $P_C$ can be related to this inhomogeneity. A droplet
model proposed by Doiron-Leyraud, {\it et al.} \cite{Doiron-Leyraud}, could
lead to an NMR signal with a dynamic signature that can observed.

\section{acknowledgments}
The work performed at UCLA and FSU was supported by the National Science
Foundation under grant numbers DMR-0203806 (SB) and DMR-0203214 (ZF).
Work at Los Alamos National Laboratory was performed under the auspices
of the U.S. Department of Energy. The authors were benefited from discussions
with W.G. Clark, 
S. Kivelson, W. Hines, and J. Budnick.

\end{document}